\documentclass{ws-procs9x6}

\newtheorem{note}{Note}[section]

\newtheorem{comment}{Comment}[section]

\def\binomial#1#2{\left(\,\begin{matrix} #1 \cr #2\end{matrix}\,\right)}

\begin{document}

\title{One-parameter groups and combinatorial physics}

\author{G\'erard Duchamp}

\address{Universit\'{e} de Rouen\\
Laboratoire d'Informatique Fondamentale et Appliqu\'{e}e de Rouen
\\ 76821 Mont-Saint Aignan, France}

\author{Karol A. Penson}

\address{Universit\'{e} Pierre et Marie Curie,\\
Laboratoire   de  Physique   Th\'{e}orique  des  Liquides, CNRS UMR 7600\\
Tour 16, $5^{i\grave{e}me}$ \'{e}tage, 4, place Jussieu, F 75252 Paris Cedex 05,
France}

\author{Allan I. Solomon}

\address{The Open University\\
 Physics and Astronomy Department\\
Milton Keynes MK7 6AA, United Kingdom}

\author{Andrej Horzela}

\address{H. Niewodnicza{\'n}ski Institute of Nuclear Physics,
Polish Academy of Sciences\\
Department of Theoretical Physics\\
ul. Radzikowskiego 152, PL 31-342 Krak{\'o}w, Poland}

\author{Pawel Blasiak}

\address{H. Niewodnicza{\'n}ski Institute of Nuclear Physics,
Polish Academy of Sciences\\
Department of Theoretical Physics\\
ul. Radzikowskiego 152, PL 31-342 Krak{\'o}w, Poland\\
\&\\
Universit\'{e} Pierre et Marie Curie,\\
Laboratoire   de  Physique   Th\'{e}orique  des  Liquides, CNRS UMR 7600\\
Tour 16, $5^{i\grave{e}me}$ \'{e}tage, 4, place Jussieu, F 75252 Paris Cedex 05,
France}

\maketitle

\abstracts{In this communication, we consider the {\em normal ordering} of operators
of the type
$$\Omega=\sum_{\alpha+\beta=e+1}c_{\alpha,\beta}(a^+)^\alpha a(a^+)^\beta,$$ where
$a$ ({\em resp.} $a^{+}$)is a boson annihilation ({\em resp.} creation) operator;
these satisfy $[a,a^{+}]\equiv a a^{+}-a^{+}a=1$, and for the purposes of this note
may be thought of as $a\equiv d/dx$ and $a^{+}\equiv x$.  We discuss the integration
of the one-parameter groups $e^{\lambda\Omega}$ and their combinatorial by-products.
In particular we show how these groups can be realized as groups of substitutions with
prefunctions.}

\section{Introduction}\label{Sw}

This text is the continuation of a series of works on the combinatorial and analytic aspects
of normal forms of boson
strings[\cite{BPS1,BPS2,BPS3,BPS4,BDHPS-tbp,Duchamp,MPBS-tbp,Krakow,Myczkowce,PS,KPS,SBDHP}].
\\
Let $w\in \{a,a^+\}^*$ be a word in the letters  $\{a,a^+\}$ (i.e. a boson string),
and define
(as in Blasiak, Penson and Solomon\cite{BPS1,BPS2,BPS3,BPS4})
by $r,s$ and $e$, respectively $|w|_{a^+}$ (the number of creation operators),
$|w|_a$ (the number of annihilation operators) and $r-s$ (the excess),
then the normal form of $w^n$ is

\begin{equation}\label{normal_form1}
\mathcal{N}(w^n)=(a^+)^{ne}\left(\sum_{k=0}^\infty S_w(n,k) (a^+)^ka^k\right)
\end{equation}
when $e$ is positive (i.e. there are more creation than annihilation operators).\\
In the opposite case (i.e. there are more annihilations than creations) the normal form of $w^n$ is
\begin{equation}\label{normal_form2}
\mathcal{N}(w^n)=\left(\sum_{k=0}^\infty S_w(n,k) (a^+)^ka^k\right)(a)^{n|e|}
\end{equation}
in each case, the coefficients $S_w$ are defined by the corresponding equation (\ref{normal_form1} and \ref{normal_form2}).\\
Now, for any boson string $u$ one has
\begin{equation}\label{lead_term}
\mathcal{N}(u)=(a^+)^{|u|_{a^+}}a^{|u|_a}+ \sum_{|v|<|u|}\lambda_v v.
\end{equation}
It has been observed [\cite{Leeuwen}] that the numbers $\lambda_v$ are {\em rook numbers}.

Consider, as examples, the upper-left corner of the following (doubly infinite) matrices.

\smallskip
For $w=a^+a$, one gets the usual matrix of Stirling numbers of the second kind.
\begin{equation}
\left\lceil
{\begin{array}{rrrrrrrr}
1 & 0 & 0 & 0 & 0 & 0 & 0 &\cdots\\
0 & 1 & 0 & 0 & 0 & 0 & 0 &\cdots\\
0 & 1 & 1 & 0 & 0 & 0 & 0 &\cdots\\
0 & 1 & 3 & 1 & 0 & 0 & 0 &\cdots\\
0 & 1 & 7 & 6 & 1 & 0 & 0 &\cdots\\
0 & 1 & 15 & 25 & 10 & 1 & 0 &\cdots\\
0 & 1 & 31 & 90 & 65 & 15 & 1&\cdots\\
\vdots & \vdots & \vdots  & \vdots  & \vdots  & \vdots  & \vdots &\ddots\\
\end{array}}
 \right.
\end{equation}

\smallskip
For $w=a^+aa^+$, we have
\begin{equation}
\left\lceil
{\begin{array}{rrrrrrrr}
1 & 0 & 0 & 0 & 0 & 0 & 0 &\cdots\\
1 & 1 & 0 & 0 & 0 & 0 & 0 &\cdots\\
2 & 4 & 1 & 0 & 0 & 0 & 0 &\cdots\\
6 & 18 & 9 & 1 & 0 & 0 & 0 &\cdots\\
24 & 96 & 72 & 16 & 1 & 0 & 0 &\cdots\\
120 & 600 & 600 & 200 & 25 & 1 & 0 &\cdots\\
720 & 4320 & 5400 & 2400 & 450 & 36 & 1&\cdots\\
\vdots & \vdots & \vdots  & \vdots  & \vdots  & \vdots  & \vdots &\ddots\\
\end{array}}
\right.
\end{equation}

\smallskip
For $w=a^+aaa^+a^+$, one gets
\begin{equation}
\left\lceil
{\begin{array}{rrrrrrrrrr}
1 & 0 & 0 & 0 & 0 & 0 & 0 & 0 & 0 & \cdots\\
2 & 4 & 1 & 0 & 0 & 0 & 0 & 0 & 0 &\cdots\\
12 & 60 & 54 & 14 & 1 & 0 & 0 & 0 & 0 &\cdots\\
144 & 1296 & 2232 & 1296 & 306 & 30 & 1 & 0 & 0 &\cdots\\
2880 & 40320 & 109440 & 105120 & 45000 & 9504 & 1016 & 52 & 1 &\cdots\\
\vdots & \vdots & \vdots  & \vdots  & \vdots  & \vdots  & \vdots & \vdots & \vdots &\ddots\\
\end{array}}
\right.
\end{equation}

\begin{remark} In each case, the matrix $S_w$ is of  staircase form and the ``step'' depends on the number of $a$'s in the word $w$.
More precisely, due to equation (\ref{lead_term}) one can prove that each row ends
with a 'one' in the cell  $(n,nr)$, where $r=|w|_a$ and we number the entries from
$(0,0)$. Thus all the matrices are row-finite and unitriangular iff $r=1$, which case
will be of special interest in the following. Moreover, the first column is
$(1,0,0\cdots, 0,\cdots, 0,\cdots)$ iff $w$ ends with an $a$ (this means that
$\mathcal{N}(w^n)$ has no  constant term for all $n>0$).
\end{remark}

In this communication, we concentrate on boson strings and more generally (homogeneous) boson operators involving only one ``$a$''. We
will see that this case is closely related to one-parameter substitution groups and their conjugates.\\
The structure of the paper is the following.\\
 In section 2 we define the framework for our transformation matrices
(spaces, topology and decomposition), then we concentrate on the {\em Riordan subgroup} (i.e. transformations which are substitutions with
prefunctions) and adapt the classical theory ({\em Sheffer condition}) to the present context.
In section 3 we analyse {\em unipotent} transformations (Lie group structure and combinatorial examples).
The divisibility property of the group of unipotent transformations tells us that every transformation is embedded in a
one-paramater group. This will be analysed in section 4 from the formal and analytic points of view.
Section 5 is devoted to some concluding remarks and further interesting possibilities.

\section{The algebra $\mathcal{L}(\mathbf{C}^\mathbf{N})$ of sequence transformations}

Let $\mathbf{C}^\mathbf{N}$ be the vector space of all complex sequences, endowed with the Frechet product topology.
It is easy to check that the algebra $\mathcal{L}(\mathbf{C}^\mathbf{N})$ of all continuous operators
$\mathbf{C}^\mathbf{N} \rightarrow \mathbf{C}^\mathbf{N}$ is the space of {\it row-finite} matrices with complex coefficients. Such a matrix
$M$ is indexed by $\mathbf{N}\times \mathbf{N}$ and has the property that, for every fixed row index $n$, the sequence $\left(M(n,k)\right)_{k\geq 0}$ has finite support.  For a sequence $A=(a_n)_{n\geq 0}$,
the transformed sequence $B=MA$ is given by $B=(b_n)_{n\geq 0}$ with
\begin{equation}
b_n=\sum_{k\geq 0} M(n,k)a_k
\end{equation}
Remark that the combinatorial coefficients $S_w$ defined above are indeed row-finite matrices.

We may associate a univariate series with a given sequence $(a_n)_{n\in \mathbf{N}},$ using a sequence
of prescribed (non-zero) denominators $(d_n)_{n\in \mathbf{N}}$ ,
as follows:
\begin{equation}
\sum_{n\geq 0}a_n \frac{z^n}{d_n}.
\end{equation}
For example, with $d_n=1$, we get the ordinary generating functions (OGF), with $d_n=n!$, we get the exponential generating functions
(EGF) and with $d_n=(n!)^2$, the doubly exponential generating functions (DEGF) and so on. Thus, once the denominators have been
chosen, to every (linear continuous) transformation of generating functions, one can associate a corresponding matrix.

\smallskip
The algebra $\mathcal{L}(\mathbf{C}^\mathbf{N})$ possesses many interesting
subalgebras and groups, such  as the algebra of lower triangular transformations
$\mathcal{T}(\mathbf{N},\mathbf{C})$, the group
$\mathcal{T}_{inv}(\mathbf{N},\mathbf{C})$ of invertible elements of the latter (which
is the set of infinite lower triangular matrices with non-zero elements on the
diagonal), the subgroup of unipotent transformations
$\mathcal{UT}(\mathbf{N},\mathbf{C})$ (i.e. the set of infinite lower triangular
matrices with elements on the diagonal all equal to $1$) and its Lie algebra
$\mathcal{NT}(\mathbf{N},\mathbf{C})$, the algebra of locally nilpotent
transformations (with zeroes on the diagonal). One has the inclusions (with
$\mathcal{D}_{inv}(\mathbf{N},\mathbf{C})$, the set of invertible diagonal matrices)
\begin{eqnarray}
\mathcal{UT}(\mathbf{N},\mathbf{C})\subset \mathcal{T}_{inv}(\mathbf{N},\mathbf{C}) \subset \mathcal{T}(\mathbf{N},\mathbf{C})
\subset \mathcal{L}(\mathbf{C}^\mathbf{N})\cr
\mathcal{D}_{inv}(\mathbf{N},\mathbf{C}) \subset \mathcal{T}_{inv}(\mathbf{N},\mathbf{C})\ {\rm and}\
\mathcal{NT}(\mathbf{N},\mathbf{C}) \subset \mathcal{L}(\mathbf{C}^\mathbf{N}).
\end{eqnarray}

We remark that
$\mathcal{T}_{inv}(\mathbf{N},\mathbf{C})= \mathcal{D}_{inv}(\mathbf{N},\mathbf{C}) \bowtie \mathcal{UT}(\mathbf{N},\mathbf{C})$
because $\mathcal{UT}$ is normalized by $\mathcal{D}_{inv}$ and $\mathcal{T}_{inv}=\mathcal{D}_{inv}.\mathcal{UT}$ (every invertible
transformation is the product of its diagonal by a unipotent transformation).

\smallskip
We now examine an important class of transformations of $\mathcal{T}$ as well as their diagonals: the substitutions with prefunctions.

\subsection{Substitutions with prefunctions}

Let $(d_n)_{n\geq 0}$ bet a fixed set of denominators. We consider, for a generating function $f$, the transformation
\begin{equation} \label{subs1}
\Phi_{g,\phi}[f](x)=g(x)f(\phi(x)).
\end{equation}
The matrix of this transformation $M_{g,\phi}$ is given by the transforms of the
monomials $\frac{x^k}{d_k}$ hence
\begin{equation}\label{sheffer}
\sum_{n\geq
0}M_{g,\phi}(n,k)\frac{x^n}{d_n}=\Phi_{g,\phi}\left[\frac{x^k}{d_k}\right]=g(x)\frac{\phi(x)^k}{d_k}.
\end{equation}

If $g,\phi\not=0$ (otherwise the transformation  is trivial), we can write

\begin{equation}\label{subs1}
g(x)=a_l\frac{x^l}{d_l} +\sum_{r>l}a_r\frac{x^r}{d_r},\ \phi(x)=
\alpha_m\frac{x^m}{d_m} +\sum_{s>m}\alpha_s\frac{x^s}{d_s}
\end{equation}

with $a_l,\alpha_m\not=0$ and then, by (\ref{sheffer},\ref{subs1})
\begin{equation}
\Phi_{g,\phi}\left[\frac{x^k}{d_k}\right]=a_l(\alpha_m)^k\frac{x^{l+mk}}{d_ld_m^kd_k}+\sum_{t>l+mk}b_t\frac{x^t}{d_t}.
\end{equation}

One then has

\begin{equation}
M_{g,\phi}\ {\rm is\ row-finite}\Longleftrightarrow \phi\ {\rm has\ no\ constant\ term}
\end{equation}

and in this case it is always lower triangular.

\smallskip
The converse is true in the following sense. Let $T\in \mathcal{L}(\mathbf{C}^{\mathbf{N}})$ be a matrix with non-zero two first
columns and suppose that the first index $n$ such that $T(n,k)\not=0$ is less for $k=0$ than $k=1$ (which is, from (\ref{sheffer}) the case
when $T=M_{g,\phi}$). Set
\begin{equation}
g(x):=d_0 \sum_{n\geq 0}T(n,0)\frac{x^n}{d_n};\ \phi(x):=\frac{d_1}{g(x)} \sum_{n\geq 0}T(n,1)\frac{x^n}{d_n}
\end{equation}
then $T=M_{g,\phi}$ iff, for all $k$,
\begin{equation}
\sum_{n\geq 0}T(n,k)\frac{x^n}{d_n}=g(x)\frac{\phi(x)^k}{d_k}.
\end{equation}

\begin{remark} Eq. (\ref{sheffer}) is called the {\em Sheffer condition} (see [\cite{Roman,SBDHP,Stanley,Wilf}]) and, for EGF ($d_n=n!$) it
amounts to stating that
\begin{equation}
\sum_{n,k\geq 0}T(n,k)\frac{x^n}{n!}y^k=g(x)e^{y\phi(x)}.
\end{equation}
\end{remark}

From now on, we will suppose that $\phi$ has no constant term ($\alpha_0=0$).\\
Moreover $M_{g,\phi}\in \mathcal{T}_{inv}$ if and only if  $a_0,\alpha_1\not=0$ and
then the diagonal term with address $(n,n)$ is
$\frac{a_0}{d_0}\left(\frac{\alpha_1}{d_1}\right)^n$. We get
\begin{equation}
M_{g,\phi}\in \mathcal{UT}\Longleftrightarrow \frac{a_0}{d_0}=\frac{\alpha_1}{d_1}=1.
\end{equation}
In particular for the EGF and the OGF, we have the condition that
\begin{equation}
g(x)=1+{\rm higher\ terms}\ and\ \phi(x)=x+{\rm higher\ terms}.
\end{equation}

\begin{note}\label{Riordan}
In classical combinatorics (for OGF and EGF), the matrices $M_{g,\phi}(n,k)$ are known as {\it Riordan matrices}
(see [\cite{Roman,Shapiro}] for example).
\end{note}

\section{Unipotent transformations}
\subsection{Lie group structure}

We first remark that $n\times n$ truncations (i.e. taking the $[0..n]\times [0..n]$
submatrix of a matrix) are algebra morphisms
\begin{equation}
\tau_n : \mathcal{T}(\mathbf{N},\mathbf{C})\rightarrow \mathcal{M}([0..n]\times [0..n],\mathbf{C}).
\end{equation}
We can endow  $\mathcal{T}(\mathbf{N},\mathbf{C})$ with the Frechet topology given by
these morphisms. We will not develop this point in detail here, but this topology is
metrisable and given by the following convergence criterion:

\smallskip
\begin{center}
{\it a sequence $(M_k)$ of matrices in $\mathcal{T}(\mathbf{N},\mathbf{C})$ converges iff\\
for\ all\ fixed\ $n\in \mathbf{N}$\\
the sequence of truncated matrices $\left(\tau_n(M_k)\right)$ converges.}
\end{center}

\smallskip
This topology is compatible with the $\mathbf{C}$-algebra structure of  $\mathcal{T}(\mathbf{N},\mathbf{C})$.

\smallskip
The two maps $\exp : \mathcal{NT}(\mathbf{N},\mathbf{C})\rightarrow \mathcal{UT}(\mathbf{N},\mathbf{C})$ and
$\log : \mathcal{UT}(\mathbf{N},\mathbf{C})\rightarrow \mathcal{NT}(\mathbf{N},\mathbf{C})$ are continous and mutually
inverse.

\subsection{Examples}
\subsubsection{Provided by the exponential formula}

The ``classical exponential formula'' [\cite{Flajolet,Joyal,Stanley}] tells us the following:
Consider a class
\footnote{Closed under relabelling (of the vertices), disjoint union, and taking connected components.}
 of finite labelled graphs
$\mathcal{C}$. Denote by  $\mathcal{C}^c$ the subclass of connected graphs in
$\mathcal{C}$. Then  the exponential generating series of $\mathcal{C}$ and
$\mathcal{C}^c$ are related as follows:
\begin{equation}
EGF(\mathcal{C})=e^{EGF(\mathcal{C}^c)}.
\end{equation}
The following examples give us some insight into  why combinatorial matrices of the
type:
\begin{center}
\small{$T(n,k)=$ {\it Number of graphs of  $\mathcal{C}$ on $n$ vertices having $k$
connected components}}
\end{center}
give rise to substitution transformations.

\begin{example} {\bf Stirling numbers.}\\
Consider the class of graphs of equivalence relations. Then using the statistics
$x^{\textrm{(number of points)}}y^{\textrm{(number of connected components)}}$
we get

\begin{eqnarray}
\sum_{n,k\geq 0}S(n,k) \frac{x^n}{n!} y^k=\cr
\hspace{-15mm}\sum_{\textrm {all equivalence graphs } \Gamma}
\frac{x^{\textrm{(number of points of }\Gamma)}} {\textrm{\scriptsize (number of points of $\Gamma$)}!}
y^{\textrm{(number of connected components of $\Gamma$)}}=\cr
\exp\left(
\sum_{\Gamma\ connected}
\frac{x^{\textrm{(number of vertices of }\Gamma)}} {\textrm{\scriptsize (number of points of $\Gamma$)}!}
y^{\textrm{(number of connected components of $\Gamma$)}}
\right)=\cr
\exp\left(
\sum_{n\geq 1} y \frac{x^n}{n!}\right)=e^{y(e^x-1)}\hspace{1cm}
\end{eqnarray}
\end{example}

we will see that the transformation associated with the matrix $S(n,k)$ is
 $f\rightarrow f(e^x-1)$.

\begin{example} {\bf Idempotent numbers.}\\
We consider the graphs of endofunctions\footnote{Functions from a finite set into itself.}. Then, using the statistics
$x^{\textrm{(number of points of the set)}}y^{\textrm{(number of connected components of the graph)}}$ and denoting by
$I(n,k)$ the number of endofunctions of a given set with $n$ elements having $k$ connected components, we get

\begin{eqnarray}
\sum_{n,k\geq 0}I(n,k) \frac{x^n}{n!} y^k=\cr
\hspace{-15mm}\sum_{\textrm {all graphs of endofunctions }\Gamma}
\frac{x^{\textrm{(number of vertices of }\Gamma)}} {\textrm{\scriptsize (number of vertices of $\Gamma$)}!}
y^{\textrm{(number of connected components of $\Gamma$)}}=\cr
\exp\left(
\sum_{\Gamma\ connected}
\frac{x^{\textrm{(number of vertices of }\Gamma)}} {\textrm{\scriptsize (number of vertices of $\Gamma$)}!}
y^{\textrm{(number of connected components of $\Gamma$)}}
\right)=\cr
\exp\left(
\sum_{n\geq 1} y \frac{nx^n}{n!}\right)=e^{yxe^x}\hspace{1cm}.
\end{eqnarray}

Corresponding to  these numbers we get the (doubly) infinite matrix

\begin{equation}
\left\lceil
{\begin{array}{rrrrrrrr}
1 & 0 & 0 & 0 & 0 & 0 & 0 &\cdots\\
0 & 1 & 0 & 0 & 0 & 0 & 0 &\cdots\\
0 & 2 & 1 & 0 & 0 & 0 & 0 &\cdots\\
0 & 3 & 6 & 1 & 0 & 0 & 0 &\cdots\\
0 & 4 & 24 & 12 & 1 & 0 & 0 &\cdots\\
0 & 5 & 80 & 90 & 20 & 1 & 0 &\cdots\\
0 & 6 & 240 & 540 & 240 & 30 & 1&\cdots\\
\vdots & \vdots & \vdots  & \vdots  & \vdots  & \vdots  & \vdots &\ddots\\
\end{array}}
\right.
\end{equation}

\end{example}

and we will see that the transformation associated with this matrix is $f\rightarrow ~f(xe^x)$

\subsubsection{Normal ordering powers of boson strings}\label{no}

To get unipotent matrices, one has to consider boson strings with only one annihilation operator.
In the introduction, we have given examples with $a^+a,a^+aa^+$ (the matrix of the third string, $a^+aaa^+a^+$,
with two annihilators, is not unipotent).
Such a string is then $w=(a^+)^{r-p}a(a^+)^p$ and we will see shortly  that

\begin{itemize}
\item if $p=0$, $S_w(n,k)$ is the matrix of a unipotent substitution
\item if $p>0$, $S_w(n,k)$ is the matrix of a unipotent substitution with prefunction
\end{itemize}

To cope with the matrices coming from the normal ordering of powers of boson strings we have to
make a small detour to analysis and formal groups.

\section{One-parameter subgroups of $UT(\mathbf{N},\mathbf{C})$}

\subsection{Exponential of elements of $NT(\mathbf{N},\mathbf{C})$}\label{tp}

Let $M=I+N\in UT(\mathbf{N},\mathbf{C})$ ($I=I_\mathbf{N}$ is the indentity matrix). One has
\begin{equation}\label{gexp}
M^t=\sum_{k\geq 0}\binomial{t}{k}N^k
\end{equation}
where $\binomial{t}{k}$ is the generalized binomial coefficient defined by
\begin{equation}
\binomial{t}{k}=\frac{t(t-1)\cdots (t-k+1)}{k!}.
\end{equation}
One can see that, for $k\leq n$, due to the local nilpotency of $N$, the matrix coefficient $M^t(n,k)$ is well defined and, in fact,
a polynomial of degree $n-k$ in $t$ (for $k>n$, this coefficient is $0$). We have the additive property $M^{t_1+t_2}=M^{t_1}M^{t_2}$
and the correspondence
$t\rightarrow M^t$ is continuous. Conversely, let $t\rightarrow M_t$ be
a continous local one-parameter group in $UT(\mathbf{N},\mathbf{C})$; that is, for some real $\epsilon>0$

\begin{equation}
|t_1|\textrm{ and }|t_2|<\epsilon \Longrightarrow M_{t_1}M_{t_2}=M_{t_1+t_2}
\end{equation}

then there exists a unique matrix $H\in NT(\mathbf{N},\mathbf{C})$ such that $M_t=exp(tH)$. (This may be proved using the projections $\tau_n$
and the classical theorem about continuous one-parameter subgroups of Lie groups, see [\cite{Kaplansky}], for example).\\
When  $M_t=M^t$ is defined by formula (\ref{gexp}) we have\\
 $H=log(I+N)=\sum_{k\geq 1}\frac{(-1)^{k-1}}{k}N^k$.

\smallskip
The mapping $t\rightarrow M^t$ will be called a {\it one parameter group} of $UT(\mathbf{N},\mathbf{C})$.

\bigskip
\begin{proposition} Let $M$ be the matrix of a substitution with prefunction; then  so is $M^t$ for all $t\in \mathbf{C}$.
\end{proposition}

\smallskip
 The proof will be given in a forthcoming paper and uses the fact that ``to be the matrix of a substitution with prefunction'' is a property
of polynomial type. But, using composition, it is straightforward to see that $M^t$ is the matrix of a substitution with prefunction for all
$t\in \mathbf{N}$. Thus, using a ``Zariski-type'' argument, we get the result that the property is true for all $t\in \mathbf{C}$.

\subsection{Link with local Lie groups : Straightening vector fields on the line}\label{paradigm}

We first  treat the case  $p=0$ of subsection (\ref{no}).  The string $(a^+)^ra$
corrresponds, in the Bargmann-Fock representation, to the vector field
$x^r\frac{d}{dx}$ defined on the whole line.\\
Now, we can try (at least locally) to straighten this vector field by a diffeomorphism $u$  to get the constant vector field (this procedure
has been introduced by G. Goldin in the context of current algebras[\cite{GG}]). As the one-parameter group generated by a constant
field is a shift, the one-parameter (local) group of transformations will be, on a suitable domain
\begin{equation}
U_\lambda[f](x)=f\left(u^{-1}\left(u(x)+\lambda\right)\right).
\end{equation}
Now, we know from section (\ref{tp}) that, if two one-parameter groups have the same tangent vector at the origin, then they coincide
{\it (tangent paradigm)}.\\
Direct computation gives this tangent vector :

\begin{equation}
\left.\frac{d}{d\lambda}\right|_{\lambda=0}f\left(u^{-1}\left(u(x)+\lambda\right)\right)=\frac{1}{u'(x)}f'(x)
\end{equation}

and so the local one-parameter group $U_\lambda$ has $\frac{1}{u'(x)}\frac{d}{dx}$ as tangent vector field.\\
Here, we have to solve $\frac{1}{u'(x)}=x^r$ in order to get the diffeomorphism $u$.\\
In the case $r\not=1$, we have (with $\mathcal{D}=]0,+\infty[$ as domain)
\begin{equation}
u(x)=\frac{x^{1-r}}{1-r}=y;\ u^{-1}(y)=\left((1-r)y\right)^{\frac{1}{1-r}}
\end{equation}
and
\begin{equation}
e^{\lambda x^r\frac{d}{dx}}[f](x)=f\left(
\frac{x}{
\left(1-\lambda(r-1)x^{r-1} \right)^{ \frac{1}{r-1} }
}
\right)
\end{equation}

The substitution factor $s_{\lambda}(x)=\frac{x}{\left(1-\lambda(r-1)x^{r-1} \right)^{ \frac{1}{r-1} }}$ has been
already obtained by other means in [\cite{BPS1}]. The computation is similar for the case when $r=1$ and, for this case,
we get
\begin{equation}
e^{\lambda x\frac{d}{dx}}[f](x)=f\left(e^\lambda x\right)
\end{equation}
with $s_{\lambda}(x)=e^\lambda x$ as substitution factor.\\
These first examples are summarized in the following table

\begin{center}
\begin{tabular}{c|c|c}
$r$    & $s_\lambda(x)$   & Name\\
\hline
$0$     & $x+\lambda$           & Shift\\
\hline
$1$ & $e^\lambda x$ & Dilation\\
\hline
$2$ & $\frac{x}{1-\lambda x}$& Homography\\
\hline
$3$ & $\frac{x}{\sqrt{1-2\lambda x^2}}$& -\\
\hline
\end{tabular}
\end{center}

\begin{comment} If one uses classical analysis (i.e. convergent Taylor series), one must be careful about  the
domain where the substitutions are defined and the one-parameter groups are defined only locally.\\
For each of these examples, one can check by hand that for suitable (and small) values of  $\lambda,\mu$,
one has $s_\lambda(s_\mu(x))=s_{\lambda+\mu}(x)$ (one-parameter group property).\\
It is possible to avoid discussion of the domains by considering $\lambda,\ \mu$ as new variables
and applying the ``substitution principle''; namely by claiming that it is possible to substitute a series without constant
term in a series (in the algebra $\mathbf{C}[[x,\lambda,\mu]]$).
\end{comment}

\smallskip
Using the same method, one can start with more complicated operators.\\
Examples and substitution factors are given below

\begin{center}
\begin{tabular}{c|c|c|c}
Operator            & Substitution Factor   & Description\\
\hline
&&\\
$\left(1+(a^+)^2\right)a$   & $s_\lambda(x)=\displaystyle{\frac{xcos(\lambda)+sin(\lambda)}{cos(\lambda)-xsin(\lambda)}}$ &
One-parameter group \\
&& of homographies\\
&&\\
\hline
&&\\
$\displaystyle{\frac{\sqrt{1+(a^+)^2}}{a^+}}$ &  $s_\lambda(x)=\sqrt{x^2+2\lambda\sqrt{1+x^2}+\lambda^2}$&
Composition of quadratic\\
&& direct and inverse functions\\
&&\\
\end{tabular}
\end{center}

\subsection{Case $p>0$: another conjugacy trick and a surprising formula}

Regarding vector fields as infinitesimal generators of one-parameter groups leads to conjugacy since, if $U_\lambda$ is a one-parameter
group of transformation, so too is $VU_\lambda V^{-1}$ ($V$ being a continuous invertible operator). We could formally consider  $(a^+)^{r-p}a(a^+)^p;\ p>0$
as conjugate to $\displaystyle{\left( (a^+)^{r}a\right)}$ in this context. More generally, supposing all the terms well-defined,
if
\begin{equation*}
\Omega=u_1(x) \frac{d}{dx} u_2(x)=\frac{1}{u_2(x)}\left(u_1(x)u_2(x) \frac{d}{dx} \right) u_2(x)
\end{equation*}
then
\begin{equation}\label{shocking}
e^{\lambda\Omega}=\frac{1}{u_2(x)}\left(e^{\lambda u_1(x)u_2(x) \frac{d}{dx} } \right) u_2(x)
\end{equation}
This rather surprising formula (\ref{shocking}) may be understood as an operator equality.\\
Now, the tangent paradigm (see section \ref{paradigm}) tells us that, if we adjust this tangent vector to coincide with
$x^{r-p}\frac{d}{dx}x^p$ (recall that the original problem was the integration of the operator $\Omega=(a^+)^{r-p}a(a^+)^p;\ p>0$),
then we get the right one-parameter group. Using this ``conjugacy trick'' we get
\begin{equation}
e^{\lambda\Omega}[f](x)=\left(\frac{s_\lambda(x)}{x}\right)f(s_\lambda(x))\textrm{ with }
s_\lambda(x)=\frac{x}{\left(1-\lambda(r-1)x^{r-1} \right)^{ \frac{1}{r-1} }}
\end{equation}

\begin{remark} (i) It can be checked that, if $s_\lambda(x)$ is a substitution factor
(i.e. at least locally $s_\lambda(s_\mu(x))=s_{\lambda+\mu}(x)$)
such that $s_\lambda(0)=0$ for every $\lambda$ (which is the case in most of our examples) then the transformations defined by
$U_\lambda[f](x)=\left(\frac{s_\lambda(x)}{x}\right)f(s_\lambda(x))$ form a one-parameter (possibly local) group.
\\
(ii) It is also possible to use the ``ad'' operator (Lie adjoint) instead of ``Ad'' (conjugacy) to obtain integration formulas (see Dattoli \cite{D1}).
\end{remark}

\subsection{Characteristic series $\leftrightarrow$ one parameter group correspondence}\label{homogen}

The preceding allows us to extend integration processes to linear combinations of boson strings in the following sense.
The algebra $W_{1,\infty}$ generated by $a^+,(a^+)^{-1},a$ is graded by
\begin{equation}
\textrm{weight}(a^+)=1,\ \textrm{weight}\left((a^+)^{-1}\right)=\textrm{weight}(a)=-1
\end{equation}
and every homogeneous operator of this algebra which is of the form
\begin{equation}
\Omega=\sum_{|w|_a=1,\ \textrm{weight}(w)=e}\alpha_w w
\end{equation}
(there is only one annihilation operator in each monomial) can be integrated as above. So one would like to reconstruct the
characteristic series
\begin{equation}
\sum_{n,k}S_\Omega(n,k) \frac{x^n}{n!}y^k
\end{equation}
from  knowledge of the one-parameter subgroup $e^{\lambda \Omega}$.\\
This is the aim of the following paragraph.

For every homogeneous operator as above with $e\geq 0$, one defines the coefficients $S_\Omega(n,k)$ as in the
Introduction by

\begin{equation}
\mathcal{N}(\Omega^n)=(a^+)^{ne}\sum_{k=0}^\infty S_\Omega(n,k)(a^+)^ka^k
\end{equation}

One has the following proposition

\begin{proposition} With the definitions introduced, the following conditions are equivalent:

\begin{eqnarray}
\sum_{n,k\geq 0} S_\Omega(n,k)\frac{x^n}{n!}y^k=g(x)e^{y\phi(x)}\\
U_\lambda[f](x)=g(\lambda x^e)f\left(x\left(1+\phi(\lambda x^e)\right) \right)
\end{eqnarray}
\end{proposition}

Which solves the problem.

\section{Conclusion and remaining problems}
We have considered a class of elements of $W_{1,\infty}$ (see section \ref{homogen} for a definition) which describe some rational vector
fields on the line. For these operators, we have established a correspondence
\begin{center}
\small{One-parameter group (=integration of the field) $\leftrightarrow$\\
 Characteristic series (=coefficients of the normal ordering)}
\end{center}
We have then seen that families of combinatorial matrices give rise
through the exponential formula to substitutions.\\
Further work which remains is to study the vector fields associated with these
combinatorial matrices. Also it would be desirable to adapt this machinery to other
algebras ($q$uons, several boson modes).

\section*{Acknowledgments}
We thank Daniel Barsky, Christophe Tollu,  Fr\'ederic Toumazet and Jean-Yves Thibon for interesting remarks and observations during
earlier stages of this work.

\end{document}